# A multiferroic two-dimensional electron gas

Julien Bréhin[1], Yu Chen[2], Maria D'Antuono[2,3], Sara Varotto[1], Daniela Stornaiuolo[2,3], Cinthia Piamonteze[4], Julien Varignon[5], Marco Salluzzo[2§] and Manuel Bibes[1*]

[1] Unité Mixte de Physique, CNRS, Thales, Université Paris Saclay, 91767 Palaiseau, France
[2] CNR-SPIN, Complesso Monte S. Angelo - Via Cinthia, 80126 Napoli, Italy
[3] University of Naples "Federico II", Complesso Monte S. Angelo - Via Cinthia, 80126 Napoli, Italy
[4] Swiss Light Source, Paul Scherrer Institut, 5232 Villigen PSI, Switzerland
[5] CRISMAT, CNRS UMR 6508, ENSICAEN, Normandie Université, 6 boulevard Maréchal Juin, 14050 Caen Cedex 4, France

Multiferroics are compounds in which at least two ferroic orders coexist – typically (anti)ferromagnetism and ferroelectricity (*1*). While magnetic order can arise in both insulating and conducting compounds, ferroelectricity is in principle not allowed in metals (*2*) although a few two-dimensional (semi)metals were reported to behave as ferroelectrics (*3–5*). Yet, the combination with magnetic order to realize multiferroic metals remains elusive. Here, by combining x-ray spectroscopy and magnetotransport, we show the coexistence of ferroelectricity and magnetism in an oxide-based two-dimensional electron gas (2DEG). The data evidence a non-volatile switching of the polar displacements and of the anomalous Hall effect by the polarization direction, demonstrating a magnetoelectric coupling. Our findings provide new opportunities in quantum matter stemming from the interplay between ferroelectricity, ferromagnetism and Rashba spin-orbit coupling in a 2DEG.

[§] marco.salluzzo@spin.cnr.it
[*] manuel.bibes@cnrs-thales.fr

Nearly 70 years ago, Anderson and Blount introduced the concept of ferroelectric metals (*2*) in which ferroelectricity and conductivity coexist in the same phase, which *a priori* seems physically impossible because itinerant electrons screen electrostatic forces between ions. More recently, LiOsO$_3$ was confirmed as the first unambiguous ferroelectric-like metal (*6*) and ferroelectricity was later evidenced in flakes of conducting WTe$_2$ (*3*, *4*). Another approach to generate a ferroelectric metal is based on the introduction of charge carriers in a bulk ferroelectric such as BaTiO$_3$ (*7*) or Ca-substituted SrTiO$_3$ (*8*) (Ca-STO). Signatures of the structural transitions associated with the ferroelectric character are then visible in the temperature dependence of the resistivity, but in these materials the ferroelectric polarization cannot be switched. This research however inspired a related strategy aiming at the generation of a two-dimensional electron gas (2DEG) in these compounds (*5*), confining the charge carriers and thus the metallic region over a few unit cells near the surface, the rest of the ferroelectric material remaining insulating. Structural and electrostatic coupling between the insulating part and the metallic part in the ferroelectric is then harnessed to introduce switchable dipoles in the metal, thereby achieving a ferroelectric 2DEG.

While novel functionalities such a non-volatile control of spin-charge conversion were demonstrated in oxide 2DEGs formed in bulk ferroelectric SrTiO$_3$ (*9*), to date, the ferroelectric character of the 2DEG itself has not been unambiguously proven. In addition, the introduction of magnetism into ferroelectric metals – leading to the generation of multiferroic metals that would open vast perspectives for spin-orbitronics (*10*) and non-reciprocal electronics (*11*) – remains elusive.

In this paper, we merge the fields of multiferroics (*1*) and oxide interfaces (*12*) by incorporating ferroelectricity and magnetism into a 2DEG based on SrTiO$_3$ (STO). Our work echoes research on 2D van der Waals materials in which ferroelectricity (*13*, *14*), ferromagnetism (*15*, *16*) and, most recently, multiferroicity (*17*) were introduced, albeit in this latter case while keeping the material insulating. Our design strategy, sketched in Fig. 1a, circumvents the daunting challenge to stabilize ferromagnetism, ferroelectric and metallicity in a single 2D material. We propose to insert a magnetic EuTiO$_3$ (ETO) layer between LaAlO$_3$ (LAO) and STO to induce magnetism in the 2DEG (bottom left) and to replace STO by ferroelectric Ca-STO to make the 2DEG behave as a ferroelectric metal (top right). Thereby, we simultaneously induce magnetism and ferroelectricity in a metal to achieve a multiferroic 2DEG (MF-2DEG; bottom right) in which magnetism and spin-dependent transport can be controlled by ferroelectricity.

To support our design strategy, we performed density functional theory (DFT) simulations. We modelled (001) slabs of the form (STO)$_{12}$/(ETO)$_2$/(LAO)$_6$/vacuum with an AlO$_2$ termination and imposing a ferromagnetic arrangement for Eu$^{2+}$ cations (see Fig. 1b for a sketch of the system).

Previous DFT calculations on metallic ETO-based heterostructures suggest that the ground state is ferromagnetic (*18*) in agreement with experimental studies (*19*). In order to mimic the ferroelectric behaviour of Ca-STO, we applied a compressive strain of 1.6% to STO inducing a polarization along the (001) direction. Then we locked the STO atomic positions of the first four unit-cells in our slabs and considered two cases with either the polarization pointing toward ($P_{up}$) or away from ($P_{dn}$) the interface. We obtain a n-type electron gas at the interface between STO/ETO and LAO as inferred by the projected density of states (PDOS) on the upper $TiO_2$ layers reported in Fig. 1c and d. In our calculations the 2DEG thus appears as a consequence of a polar catastrophe (*20*) although we cannot exclude that oxygen vacancies may also contribute(*21*). The DOS at the Fermi level is sizeable for $P_{up}$ and smaller but finite for $P_{dn}$.

Due to the presence of the ferromagnetically ordered $Eu^{2+}$ spins, Ti cations experience a spin-dependent potential which produces asymmetries between spin-up and spin-down channels for the *d*-states. It results in a spin-polarized DOS and a net magnetic moment on Ti cations that spreads over four $TiO_2$ layers in the STO below the ETO/LAO interface. As visible in Fig. 1e, the Ti magnetic moment is maximum for the layer sandwiched between two $Eu^{2+}$ layers below the interface with LAO, but extends into the first STO unit-cells. An interesting prediction of Fig. 1e is also that the magnetic moments on Ti cations are different for $P_{up}$ and $P_{dn}$. Thus, ferroelectric switching offers a direct control of the magnetization in the MF-2DEG. Our calculations therefore predict that our 2DEGs should display a magnetoelectric coupling. This fact may originate from the different band bending induced at the STO/ETO and ETO/LAO interfaces for the two polarization states (Fig. 1b and c): $P_{up}$ ($P_{dn}$) brings a positive (negative) contribution to the electric field present at the $TiO_2$-LaO interface thereby increasing (decreasing) the band bending. Consequently, the Fermi level crosses more Ti 3*d* states and farther away from the interface for $P_{up}$ than for $P_{dn}$ state where it is limited to three $TiO_2$ layers below the interface.

One of the effects of ferroelectricity in the heterostructure is a modulation of the B-cations (Al/Ti) and oxygen ions relative displacements along the growth direction induced by the ferroelectric polarization direction, as shown in Fig. 1f. In LAO, the displacements are finite and large, due to the electric field created by the polar interface with ETO, but they are equivalent for $P_{up}$ and $P_{dn}$, within error bars. Deep into the STO, the polar displacements $\Delta z_{Ti-O}$ are also present but are different depending on the polarization state, as a consequence of ferroelectricity. Importantly, this difference persists in the 2DEG region, indicating that the 2DEG possesses two different remanent polarization states. In other words, it is ferroelectric.

To confirm this prediction, we measured the X-ray linear dichroism (XLD) at the Ti $L_{3,2}$ edge on a typical sample prepared by growing a LAO(10 uc)/ETO(2 uc) bilayer onto (001)-oriented Ca-STO single crystals using pulsed laser deposition (*19*) (see Methods and Fig. S1 for details). This element-specific technique corresponds to the difference in the X-ray absorption measured with linear vertical or horizontal polarization, i. e. parallel or perpendicular to the interface, and probes the local crystal field splitting. The data plotted in Fig. 1g are measured in total electron yield mode (TEY, probing depth ~2 nm) and thus correspond to the interfacial region including ETO and the first atomic planes of STO where the 2DEG is located. The red and blue spectra are collected at 2 K and at electrical remanence after switching the sample *in situ* into the $P_{up}$ and $P_{dn}$ state, respectively. The XLD spectra are visibly modulated by the remanent polarization direction, with the main changes seen in features B and D

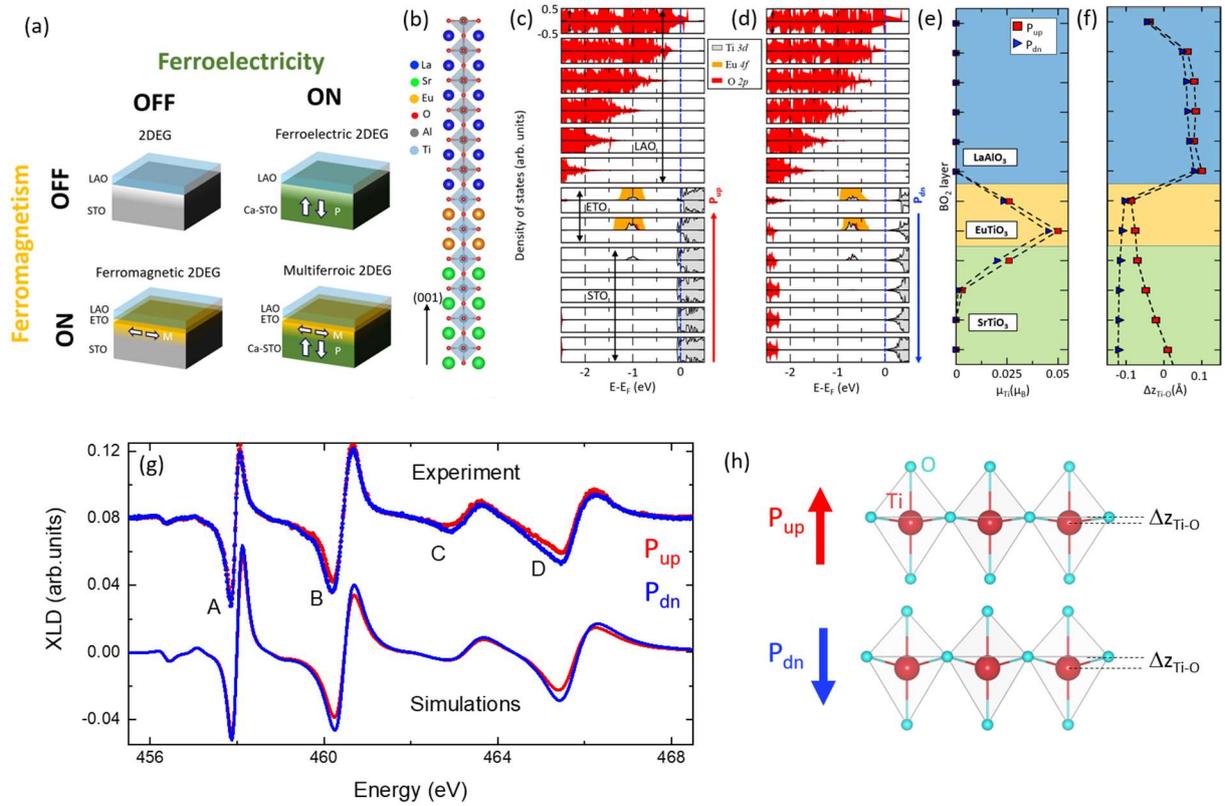

**Fig. 1. Design of a multiferroic 2DEG.** *(a) Starting from the non-magnetic, non-ferroelectric interface system LAO/STO (top left), ferroelectricity can be introduced by replacing STO by Ca-STO (top right). Ferromagnetism can be introduced by inserting a few unit cells of ETO between LAO and STO (bottom left). Combining both approaches yields a multiferroic 2DEG (bottom right). (b) Sketch of the slab configuration tested in the electronic structure simulations. (c) and (d) Layer-decomposed projected density of states on O 2p and Ti d states as a function of a polarization pointing toward (c) and outward (d) the interface. Positive and negative values stand for spin up and spin down, respectively. The Fermi level is set at 0 eV and is represented by the blue dashed line. (e) Computed magnetic moments on Ti cations for the two polar states. (f) Relative displacement of the B cation with respect to oxygen ions*

*along the growth direction. (g) X-ray linear dichroism spectra at the Ti $L_{3,2}$ edge measured at 2 K and ferroelectric remanence for $P_{up}$ and $P_{dn}$ on a LAO/ETO//Ca-STO sample (top) and atomic multiplet simulations (bottom). (h) Sketches of the $TiO_6$ octahedra configurations corresponding to the simulations in (g), evidencing different off-centered Ti-O displacements $\Delta z_{Ti-O}$ for $P_{up}$ and $P_{dn}$, consistent with the DFT calculations.*

corresponding to the Ti $e_g$ states. Atomic multiplet calculations reproducing the data show that the splitting of the $e_g$ ($t_{2g}$) states is 15 meV (10 meV) larger for $P_{dn}$ compared to $P_{up}$, consistent with the larger $\Delta z_{Ti-O}$ values computed by DFT for $P_{dn}$ compared to $P_{up}$. The corresponding $TiO_6$ octahedra configurations are sketched in Fig. 1h. These results thus indicate that the positions of the off-centered Ti ions with respect to the oxygens depend on the polarization state, thereby demonstrating the presence of switchable ferroelectric dipoles in the 2DEG and confirming its ferroelectric character.

To selectively probe the magnetic response of the elements present in the interface region, we performed further X-ray absorption spectroscopy (XAS) experiments in TEY and measured the X-ray magnetic circular dichroism (XMCD) in an in-plane magnetic field *B*. Fig. 2a-d display XAS and XMCD at 5 T for Ti and Eu. The XAS spectrum measured at the Ti $L_{3,2}$ edge (Fig. 2a) is typical of Ti in a valence state close to 4+. A small fraction of $Ti^{3+}$ (a few percent) is however expected to be present owing to the population of the Ti $t_{2g}$ states. Fig. 2b displays the normalized difference between XAS collected with right-and left-circular polarized light, i.e., the XMCD at the Ti $L_{3,2}$ edge. A clear dichroic signal of more than 2% is visible, indicating the presence of Ti magnetic moments in the sample. The shape of the XMCD is reminiscent of that measured in Ref. (*19*). Extracting the value of the Ti moment from the XMCD signal is not trivial because sum rules do not apply well to early 3*d* transition metals. However, comparison with Ref. (*19*) suggests a total moment of around 0.025 $\mu_B$/Ti. Fig. 2e shows the dependence of the XMCD signal at 457.6 eV (corresponding to the Ti $L_3$ edge) as a function of the in-plane magnetic field: a clear open hysteresis cycle is observed, consistent with a ferromagnetic behaviour probed along an easy magnetization axis. The XAS and XMCD at the Eu $M_{5,4}$ edges are presented in Fig. 2c and 2d. The XAS spectrum is compatible with $Eu^{2+}$. The large XMCD and the hysteresis loop displayed in Fig. 2f show that ETO is ferromagnetic as in electron-doped ETO (*18*, *22*), rather than antiferromagnetic as in the bulk (*23*).

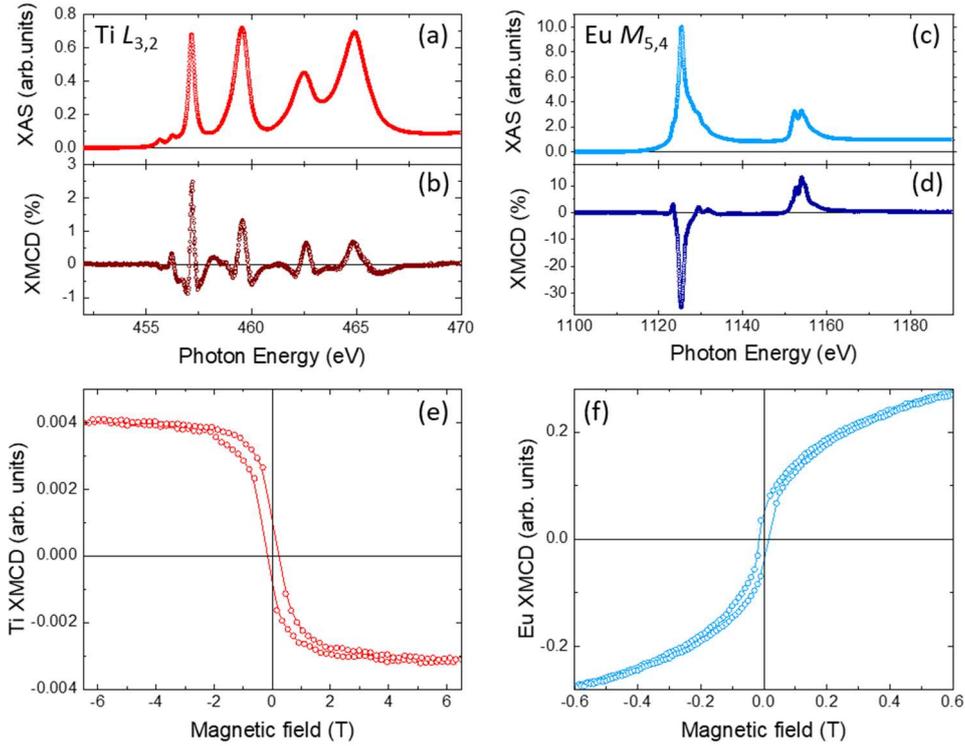

***Fig. 2. X-ray absorption spectroscopy and X-ray magnetic circular dichroism.*** *XAS spectra at the Ti $L_{3,2}$ (b) and Eu $M_{5,4}$ edges (c) with corresponding XMCD spectra (b) and (d) for a LAO/ETO//Ca-STO sample, at 2 K. Magnetic field dependence of the XMCD signal at the Ti $L_{3,2}$ (e) and Eu $M_{5,4}$ (f) edges in grazing incidence (in plane magnetic field).*

Fig. 3 illustrates the strong coupling between ferroelectricity and transport properties in our MF-2DEGs. As visible in the inset of Fig. 3a, the 2DEG displays the expected metallic temperature dependence of the sheet resistance $R_S$ upon cooling from 300 K. Near 35 K the data however show a minimum below which the resistance increases (Fig. 3a, main panel). This minimum correlates with the transition to the ferroelectric state as observed in electron-doped bulk Ca-STO (*8*) and in other 2DEGs based on Ca-STO (*5*). Fig. 3b presents polarization *P* vs $V_G$ cycles, which exhibit a clear hysteresis loop ($V_G$ is the gate voltage applied to the structure, see inset), associated with the presence of broad switching peaks in the current (*I*) *vs* $V_G$ data (see Fig. S2). The coercive field (~0.5 kV/cm) is in the range of previously reported values (*5*, *8*, *24*). The *P* vs $V_G$ cycles remain hysteretic up to 35 K. From the temperature (*T*) dependence of the remanent polarization ($P_R$) deduced from remanent polarization cycles (see Fig. S3) shown in Fig. 3c, we extract a ferroelectric Curie temperature $T_C^E$ of about 35 K, compatible with the $T_C^E$ of 1% Ca-substituted STO (*25*) and with earlier results on Ca-STO based 2DEGs (*5*, *26*).

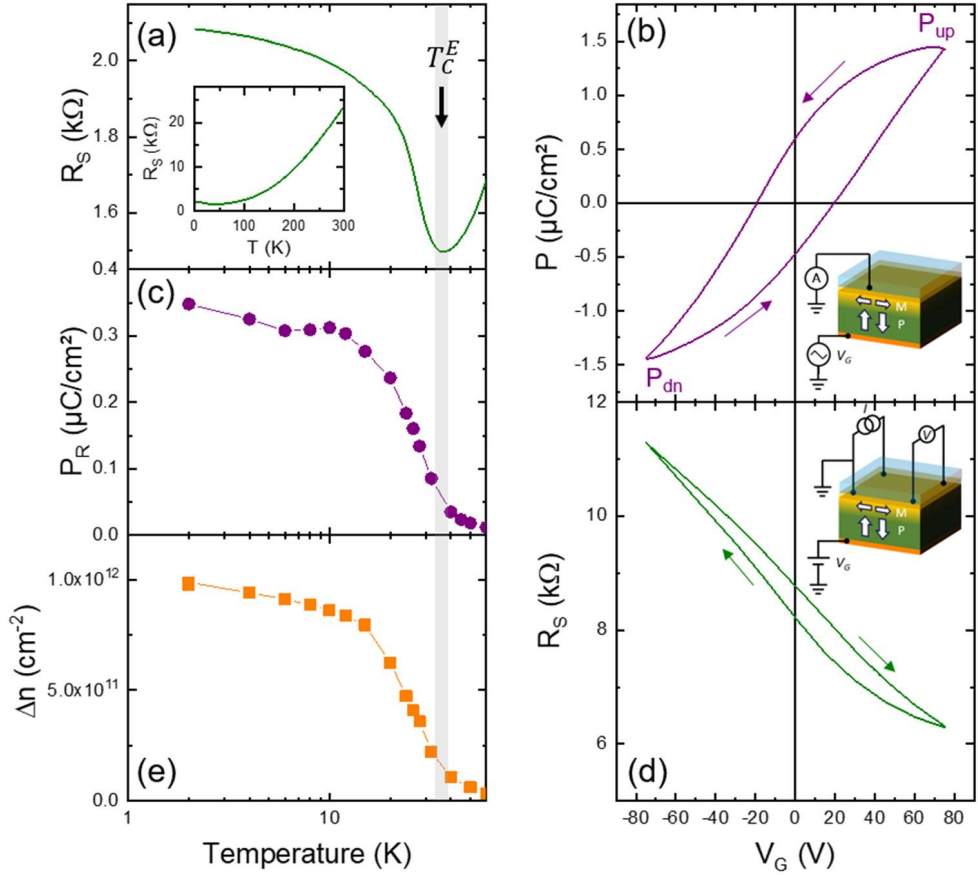

*Fig. 3. Ferroelectric and transport properties. Temperature dependence of the sheet resistance (a). (b) Polarization vs voltage loop at 2 K. (c) Temperature dependence of the remanent polarization. (d) Gate voltage dependence of the 2DEG sheet resistance at 2 K. (e) Temperature dependence of the difference in carrier density between the two ferroelectric polarization remanent states.*

Fig. 3d shows the gate dependence of the sheet resistance: after gate initialization, $R_S$ displays a clockwise hysteretic dependence with the gate voltage that is reproducible upon cycling (Fig. S4), a feature absent in equivalent 2DEGs grown on non-ferroelectric STO (*27*). $R_S$ is lower when *P* points towards the 2DEG due to electron accumulation. The data thus indicate that the electronic properties of the 2DEG are strongly modulated by ferroelectricity.

The carrier density *n* extracted from the ordinary Hall effect also depends on the remanent polarization state, $n=2.64 \times 10^{13}$ cm$^{-2}$ for P$_{up}$ and $n=2.45 \times 10^{13}$ cm$^{-2}$ for P$_{dn}$ at 2 K, which yields a difference $\Delta n=9.73 \times 10^{11}$ cm$^{-2}$. The dependence of $\Delta n$ with temperature is shown in Fig. 3e. It vanishes near 35 K, which corresponds to $T_C^E$, as expected. This $\Delta n$ corresponds to about 22% of the value of the switched remanent polarization of Ca-STO, i.e., $2P_r/e=4.38 \times 10^{12}$ cm$^{-2}$ with $P_r=0.35$ µC/cm² at 2 K.

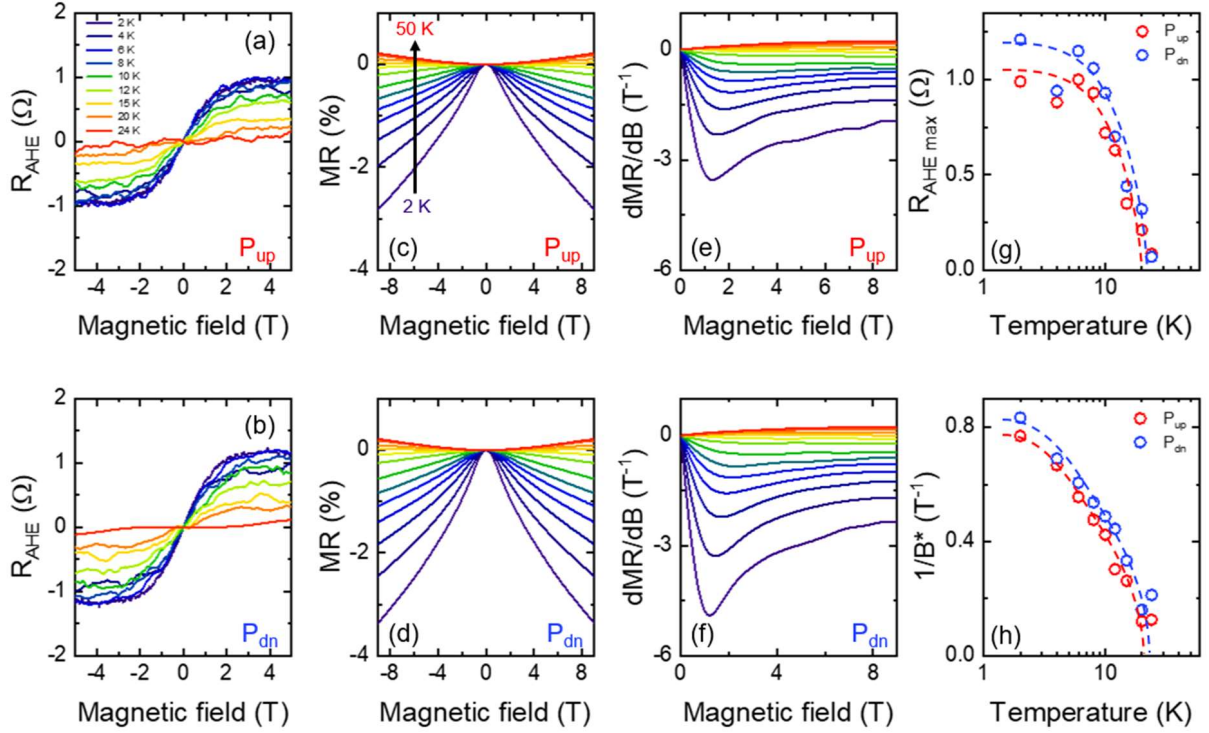

*Fig. 4. Magnetotransport properties for both ferroelectric polarization remanent states.* Anomalous Hall effect at different temperatures for $P_{up}$ (a) and $P_{dn}$ (b). Magnetoresistance (c, d) and its magnetic field derivative (e, f) at different temperatures for $P_{up}$ (c, e) and $P_{dn}$ (d, f). Temperature dependence of the maximum amplitude of the AHE (g) and of the inverse of B* (h) for $P_{up}$ and $P_{dn}$. In (g) and (h) the dashed lines are guides to the eye.

Although the Hall effect in our 2DEG is mostly linear, closer inspection reveals a nonlinear contribution at low temperature and low magnetic field, ascribed to an anomalous Hall effect (AHE) contribution due to the magnetic order of the system (see Fig. S5 and S6). In Fig. 4a and b we plot the AHE at different temperatures for $P_{up}$ and $P_{dn}$. Its shape matches that of XMCD cycles measured with the magnetic field perpendicular to the sample plane (i.e. along the hard magnetization axis), see Fig. S7. The AHE amplitude decreases with *T* and vanishes around 20 K (cf Fig. 4g) which we assign to the magnetic Curie temperature $T_C^M$. The AHE is typically higher in the $P_{dn}$ state (Fig. 4g) and its shape also depends on the ferroelectric state (see Fig. S8). These data thus point to a ferroelectric control of the AHE and of the magnetization, which indicates the presence of a magnetoelectric coupling in our 2DEGs.

Fig. 4c and d show the magnetoresistance (MR) measured with a perpendicular magnetic field as a function of temperature at positive and negative ferroelectric remanence. At high *T* (e.g. 50 K) the MR shows a small positive quasi-parabolic dependence, typical of Lorentz MR and commonly observed in oxide 2DEGs. The MR scales as $\mu^2 B^2$ with $\mu$ the electron mobility, which we estimate to about 70 cm²/Vs at 50 K, in agreement with values deduced from the sheet conductance and carrier density extracted from the ordinary Hall coefficient. As temperature decreases, the MR develops a negative field dependence that resembles what is found in magnetic systems with some degree of spin disorder (*28*): as magnetic field increases, spins rotate to align with the external magnetic field, reducing the resistance. While this low temperature perpendicular MR is anhysteretic, the MR measured with the field in the sample plane shows the expected hysteresis related to the magnetization switching shown in Fig. 2 (see Fig. S9).

At low *T*, the MR shows an inflection as a function of magnetic field, which can be better seen from the derivatives (Fig. 4e and f) and is manifested as a pronounced dip at a field *B**. The inflection may signal the approach to saturation of the magnetic lattice. Consistent with this interpretation, the dip in the derivative disappears as *T* increases. Plotting 1/*B** as function of temperature (Fig. 4h) yields a transition temperature that matches the $T_C^M$ deduced from the *T* dependence of the AHE amplitude (Fig. 4g).

From the temperature dependence of the $R_S$ vs *B* data for both ferroelectric remanent states we can plot the electroresistance ER=($R_{Pup}$-$R_{Pdn}$) at zero magnetic field and 9 T, see Fig. 5a. The electroresistance decreases with *T* and vanishes around $T_C^E$, as expected. Interestingly, ER at 0 T and 9 T almost overlap at high temperature but are different below about 20 K, which corresponds to $T_C^M$. The difference in the ER at 0 T and 9 T corresponds to an electro-magnetoresistance EMR effect (reflecting the influence of the ferroelectric remanent state on the MR, or reciprocally to the influence of magnetic field on the ER). The EMR decreases with increasing temperature and disappears when the first ferroic ordering temperature is reached, here $T_C^M$. The observation of an EMR effect is reminiscent of the behavior found in multiferroic tunnel junctions (*29*, *30*).

Fig. 5c-e illustrate the cyclability of the AHE and resistance levels upon setting different polarization states. We have consecutively applied gate voltages of +50 V, 0 V, -50 V and 0 V to respectively pole the ferroelectric upwards, set it in its positive remanent state $P_{up}$, pole the ferroelectric downwards and set it in its negative remanent state $P_{dn}$ (cf Fig. 5c). For each state we have measured the AHE and the MR, see Fig. 5d and 5e, respectively. Both the AHE and MR increase when going from more positive to more negative polarizations. Remarkably, the variations are reproducible from one cycle to the next, attesting of the robustness of the magnetoelectric coupling.

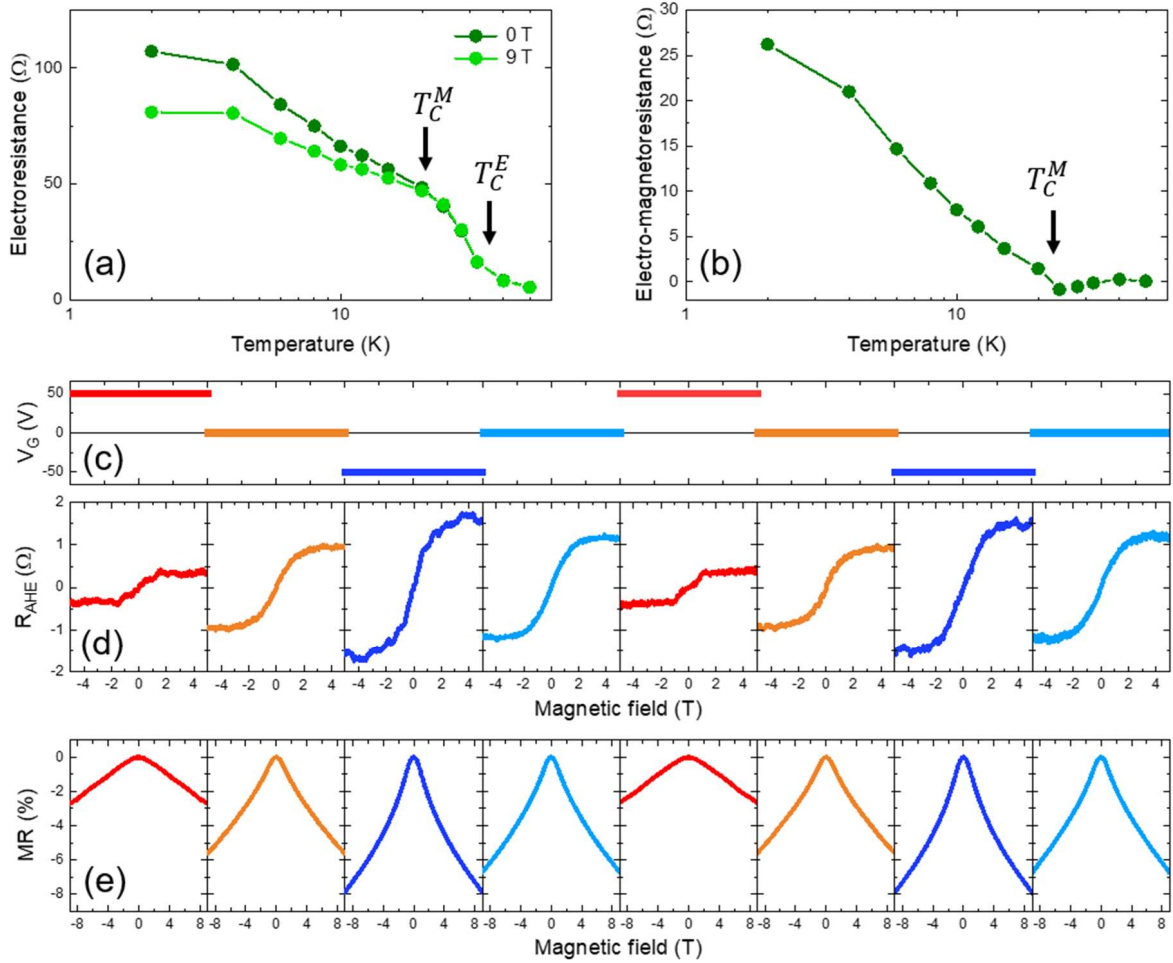

***Fig. 5. Magnetoelectric coupling.*** *(a) Temperature dependence of the electroresistance at 0 and 9 T. (b) Temperature dependence of the electro-magnetoresistance. (c) Trains of gate voltages applied to the samples. Anomalous Hall effect (d) and magnetoresistance (e) measured at 2 K at the voltages indicated in (c).*

While the multiferroic and magnetoelectric characters of the 2DEG are well supported by the data, some results deserve further investigation. For instance, although our DFT calculations suggest that *M* in the 2DEG should be higher for $P_{up}$ than for $P_{dn}$, experimentally we find the AHE amplitude to be stronger for $P_{dn}$ than for $P_{up}$. These results are not necessarily incompatible since in ETO heterostructures the AHE is known to depend non-monotonically on the saturation magnetization, arising from topological points in the band structure, and even showing a sign change with doping (*31*). From XLD data, we find that the polarization direction modifies substantially the crystal field splitting, which indicates a non-trivial P-dependence of the electronic band structure, as also suggested by DFT calculations. Another puzzling finding is the high $T_C^M \approx$ 20 K, which is higher than $T_C^M$ in strained or doped ETO (peaking around 10 K (*22*)). The different electronic structure of the ferroelectric 2DEG compared to non-ferroelectric STO 2DEGs or doped ETO (*31*), as well as enhanced correlations in this low-dimensional system, may collaborate to yield this increase.

In summary, we have designed and prepared LaAlO$_3$/EuTiO$_3$//Ca-SrTiO$_3$ heterostructures harbouring a two-dimensional electron gas that displays coupled ferroelectricity and magnetism coexisting with a metallic behaviour. Our results not only marry the fields of multiferroics and oxide interfaces but also provides a fascinating new physics playground for non-volatile spin-orbitronics (*10*) and nonreciprocal physics (*11*). They call for further theoretical calculations and experiments to address the interplay of the multiferroic behaviour with Rashba SOC. For instance, because of their intrinsic inversion symmetry breaking, MF-2DEGs are expected to display Rashba-type Dzyaloshinskii-Moriya interaction (DMI)(*32*) leading to chiral spin textures such as skyrmions (*33*) which here should be controllable by ferroelectricity.

**Acknowledgements**

The authors thank A. Barthélémy, N. Bergeal, V. Garcia, Y. Dagan and B. Kalisky for useful discussions and L.M. Vicente-Arche for the data in Supplementary Fig. 6 c and d. This project received funding from the ERC Advanced grant "FRESCO" n° 833973, the French ANR through project "CONTRABASS", the ERA-NET QUANTERA European Union's Horizon H2020 project "QUANTOX" under Grant Agreement No. 731473 and the Ministero dell'Istruzione, dell'Università e della Ricerca for the PRIN project "TOP-SPIN" (Grant No. PRIN 20177SL7HC). DFT calculations took advantages of HPC resources of CRIANN through the projects 2020005 and 2007013 and of CINES through the DARI project A0080911453.

**References**

1. M. Fiebig, T. Lottermoser, D. Meier, M. Trassin, The evolution of multiferroics. *Nat. Rev. Mater*. **1**, 16046 (2016).

2. P. W. Anderson, E. I. Blount, Symmetry Considerations on Martensitic Transformations: Ferroelectric Metals? *Phys. Rev. Lett*. **14**, 217–219 (1965).

3. Z. Fei, W. Zhao, T. A. Palomaki, B. Sun, M. K. Miller, Z. Zhao, J. Yan, X. Xu, D. H. Cobden, Ferroelectric switching of a two-dimensional metal. *Nature*. **560**, 336–339 (2018).

4. P. Sharma, F.-X. Xiang, D.-F. Shao, D. Zhang, E. Y. Tsymbal, A. R. Hamilton, J. Seidel, A room-temperature ferroelectric semimetal. *Sci. Adv.* **5**, eaax5080 (2019).

5. J. Bréhin, F. Trier, L. M. Vicente-Arche, P. Hemme, P. Noël, M. Cosset-Chéneau, J.-P. Attané, L. Vila, A. Sander, Y. Gallais, A. Sacuto, B. Dkhil, V. Garcia, S. Fusil, A. Barthélémy, M. Cazayous, M. Bibes, Switchable two-dimensional electron gas based on ferroelectric Ca:SrTiO$_3$. *Phys. Rev. Mater*. **4**, 041002 (2020).

6. Y. Shi, Y. Guo, X. Wang, A. J. Princep, D. Khalyavin, P. Manuel, Y. Michiue, A. Sato, K. Tsuda, S. Yu, M. Arai, Y. Shirako, M. Akaogi, N. Wang, K. Yamaura, A. T. Boothroyd, A ferroelectric-like structural transition in a metal. *Nature Mater*. **12**, 1024–1027 (2013).

7. T. Kolodiazhnyi, A. Petric, M. Niewczas, C. Bridges, A. Safa-Sefat, J. E. Greedan, Thermoelectric power, Hall effect, and mobility of $n$-type BaTiO$_3$. *Phys. Rev. B*. **68**, 085205 (2003).

8. C. W. Rischau, X. Lin, C. P. Grams, D. Finck, S. Harms, J. Engelmayer, T. Lorenz, Y. Gallais, B. Fauqué, J. Hemberger, K. Behnia, A ferroelectric quantum phase transition inside the superconducting dome of Sr$_{1-x}$Ca$_x$TiO$_{3-\delta}$. *Nature Phys*. **13**, 643 (2017).

9. P. Noël, F. Trier, L. M. Vicente Arche, J. Bréhin, D. C. Vaz, V. Garcia, S. Fusil, A. Barthélémy, L. Vila, M. Bibes, J.-P. Attané, Non-volatile electric control of spin–charge conversion in a SrTiO$_3$ Rashba system. *Nature*. **580**, 483–486 (2020).

10. F. Trier, P. Noël, J.-V. Kim, J.-P. Attané, L. Vila, M. Bibes, Oxide spin-orbitronics: spin–charge interconversion and topological spin textures. *Nat. Rev. Mater*. **7**, 258–274 (2022).

11. Y. Tokura, N. Nagaosa, Nonreciprocal responses from non-centrosymmetric quantum materials. *Nat. Commun*. **9**, 3740 (2018).

12. H. Y. Hwang, Y. Iwasa, M. Kawasaki, B. Keimer, N. Nagaosa, Y. Tokura, Emergent phenomena at oxide interfaces. *Nature Mater*. **11**, 103–113 (2012).

13. M. Vizner Stern, Y. Waschitz, W. Cao, I. Nevo, K. Watanabe, T. Taniguchi, E. Sela, M. Urbakh, O. Hod, M. Ben Shalom, Interfacial ferroelectricity by van der Waals sliding. *Science*. **372**, 1462–1466 (2021).

14. K. Yasuda, X. Wang, K. Watanabe, T. Taniguchi, P. Jarillo-Herrero, Stacking-engineered ferroelectricity in bilayer boron nitride. *Science*. **372**, 1458–1462 (2021).

15. C. Gong, L. Li, Z. Li, H. Ji, A. Stern, Y. Xia, T. Cao, W. Bao, C. Wang, Y. Wang, Z. Q. Qiu, R. J. Cava, S. G. Louie, J. Xia, X. Zhang, Discovery of intrinsic ferromagnetism in two-dimensional van der Waals crystals. *Nature*. **546**, 265–269 (2017).

16. B. Huang, G. Clark, E. Navarro-Moratalla, D. R. Klein, R. Cheng, K. L. Seyler, D. Zhong, E. Schmidgall, M. A. McGuire, D. H. Cobden, W. Yao, D. Xiao, P. Jarillo-Herrero, X. Xu, Layer-dependent ferromagnetism in a van der Waals crystal down to the monolayer limit. *Nature*. **546**, 270–273 (2017).

17. Q. Song, C. A. Occhialini, E. Ergeçen, B. Ilyas, D. Amoroso, P. Barone, J. Kapeghian, K. Watanabe, T. Taniguchi, A. S. Botana, S. Picozzi, N. Gedik, R. Comin, Evidence for a single-layer van der Waals multiferroic. *Nature*. **602**, 601–605 (2022).

18. Z. Gui, A. Janotti, Carrier-Density-Induced Ferromagnetism in $EuTiO_3$ Bulk and Heterostructures. *Phys. Rev. Lett.* **123**, 127201 (2019).

19. D. Stornaiuolo, C. Cantoni, G. M. D. Luca, R. D. Capua, E. D. Gennaro, G. Ghiringhelli, B. Jouault, D. Marrè, D. Massarotti, F. M. Granozio, I. Pallecchi, C. Piamonteze, S. Rusponi, F. Tafuri, M. Salluzzo, Tunable spin polarization and superconductivity in engineered oxide interfaces. *Nature Mater*. **15**, 278–284 (2016).

20. N. Nakagawa, H. Y. Hwang, D. A. Muller, Why some interfaces cannot be sharp. *Nature Mater*. **5**, 204–209 (2006).

21. L. Yu, A. Zunger, A polarity-induced defect mechanism for conductivity and magnetism at polar–nonpolar oxide interfaces. *Nat Commun*. **5**, 5118 (2014).

22. T. Katsufuji, Y. Tokura, Transport and magnetic properties of a ferromagnetic metal: $Eu_{1-x}R_xTiO_3$. *Phys. Rev. B*. **60**, R15021–R15023 (1999).

23. C.-L. Chien, S. DeBenedetti, F. D. S. Barros, Magnetic properties of $EuTiO_3$, $Eu_2TiO_4$, and $Eu_3Ti_2O_7$. *Phys. Rev. B*. **10**, 3913–3922 (1974).

24. B. S. de Lima, S. Harms, C. P. Grams, J. Hemberger, X. Lin, B. Fauque, K. Behnia, Interplay between antiferrodistortive, ferroelectric, and superconducting instabilities in $Sr_{1-x}Ca_xTiO_{3-\delta}$. *Phys. Rev. B*. **91**, 045108 (2015).

25. J. G. Bednorz, K. A. Müller, $Sr_{1-x}Ca_xTiO_3$: An XY Quantum Ferroelectric with Transition to Randomness. *Phys. Rev. Lett.* **52**, 2289–2292 (1984).

26. G. Tuvia, Y. Frenkel, P. K. Rout, I. Silber, B. Kalisky, Y. Dagan, Ferroelectric Exchange Bias Affects Interfacial Electronic States. *Adv. Mater.* **32**, 2000216 (2020).

27. J. Biscaras, S. Hurand, C. Feuillet-Palma, A. Rastogi, R. C. Budhani, N. Reyren, E. Lesne, J. Lesueur, N. Bergeal, Limit of the electrostatic doping in two-dimensional electron gases of $LaXO_3(X = Al, Ti)/SrTiO_3$. *Sci Rep*. **4**, 6788 (2015).

28. Y. Tokura, Y. Tomioka, Colossal magnetoresistive manganites. *J. Magn. Magn. Mater.* **200**, 1 (1999).

29. M. Gajek, M. Bibes, S. Fusil, K. Bouzehouane, J. Fontcuberta, A. Barthélémy, A. Fert, Tunnel junctions with multiferroic barriers. *Nature Mater*. **6**, 296–302 (2007).

30. V. Garcia, M. Bibes, L. Bocher, S. Valencia, F. Kronast, A. Crassous, X. Moya, S. Enouz-Vedrenne, A. Gloter, D. Imhoff, C. Deranlot, N. D. Mathur, S. Fusil, K. Bouzehouane, A. Barthélémy, Ferroelectric Control of Spin Polarization. *Science*. **327**, 1106 (2010).

31. K. S. Takahashi, H. Ishizuka, T. Murata, Q. Y. Wang, Y. Tokura, N. Nagaosa, M. Kawasaki, Anomalous Hall effect derived from multiple Weyl nodes in high-mobility EuTiO$_3$ films. *Sci. Adv.* **4**, eaar7880 (2018).

32. H. Imamura, P. Bruno, Y. Utsumi, Twisted exchange interaction between localized spins embedded in a one- or two-dimensional electron gas with Rashba spin-orbit coupling. *Phys. Rev. B*. **69**, 121303 (2004).

33. A. Fert, N. Reyren, V. Cros, Magnetic skyrmions: advances in physics and potential applications. *Nat. Rev. Mater*. **2**, 17031 (2017).

**Materials and Methods**

Sample preparation: LAO(n)/ETO(2)/STO heterostructures were fabricated by pulsed laser deposition (PLD) assisted by Reflection High Energy Electron Diffraction (RHEED) from sintered $Eu_2Ti_2O_7$ and crystalline LAO targets onto $TiO_2$-terminated (001) $Sr_{0.99}Ca_{0.01}TiO_3$. $TiO_2$ terminated Ca-STO substrates (from SurfaceNet GmbH) were obtained by etching the as-received single crystals in a buffered hydrofluoric acid solution (pH=5.5) and annealed in flowing $O_2$ for 2h at 1000 °C. ETO and LAO were sequentially deposited in $8\times10^{-5}$ mbar of $O_2$ at 720 °C. A Ti(20 nm)/Au(50 nm) bilayer was sputtered at as a bottom electrode the back of the substrate to apply a voltage through the substrate for polarization measurements and gate-dependent magnetotransport.

Electric polarisation measurements. In these experiments performed with a Radiant Multiferroics II tester, a triangular waveform was applied at a frequency of 10 Hz across the Ca-STO, between the Ti/Au bottom electrode (shown in orange in the sketches of Fig. 3a and 3b) and the 2DEG and the current *I* was measured in real time. The static leakage current measured in dc was less than 5 nA at 50 V (corresponding to a resistivity of 0.5 GΩ.cm). Integrating the current with time and normalizing by the sample area yields the polarisation. We used the same principle to obtain the "remanent hysteresis" curves by performing the polarization measurements four times just after applying a preset voltage pulse of different polarities: this allows to subtract the non-switching contribution from the total polarization and thus to reconstruct the four branches of the loop keeping only the remanent (switchable) polarization.

Magnetotransport. Electrical transport measurements were performed on the samples bonded by Al wires in the van der Pauw configuration in a Quantum Design Dynacool cryostat, using a standard DC technique with a Keithley 2400 Sourcemeter, while the gate voltage was applied with another Keithley 2450 Sourcemeter between the bottom electrode and the 2DEG. The sheet resistance was measured in four-wire configuration expect for Fig. 3e. Before being analysed the Hall and MR raw signals were anti-symmetrized and symmetrized with respect to magnetic field, respectively.

XAS, XMCD and XLD. The experiments were performed on the EPFL/PSI X-Treme beamline at the Swiss Light Source (*34*). We have used photon-polarization dependent x-ray absorption spectroscopy (XAS) across the Eu $M_{4,5}$ and the Ti-$L_{2,3}$ edge to probe directly the magnetic and orbital properties of Eu and Ti at the interface. Data were acquired in the total electron yield (TEY) mode, which is sensitive to the interfacial region, i.e. 2 nm from the LAO/ETO interface. The Eu $M_{4,5}$ edge and the Ti $L_{2,3}$ x-ray magnetic circular dichroism (XMCD) spectra were obtained as difference between the average of 8 and 16 (respectively) XAS spectra acquired with magnetic field parallel and antiparallel to the photon-helicity vector orientations in a sequence alternating reversal of field and polarization at each spectrum. This procedure ensures the best cancellation of spurious effects. The magnetic field dependent magnetization loops shown in Fig. 2 were obtained by measuring, at each field, the difference between the TEY intensity at the Ti-$L_3$ and $M_5$-Eu edge peaks, normalized by the intensity below the absorption edge, obtained with two different helicities (combination of polarization and field direction). The XLD spectra in Fig. 1g were obtained in grazing incidence conditions (x-ray incidence angle of 60 degrees respect the surface normal) at 2 K as the difference between the average of 8 XAS spectra acquired with linear polarization parallel (LV) and almost perpendicular (LH) to the interface. Ferroelectric polarization dependent XLD data were acquired by switching in situ the ferroelectric polarization direction with a back-gate voltage (isolated from the ground), and the 2DEG grounded. To verify the effective switching of the polarization during the experiment, several P vs V loops were acquired before the acquisition of each XLD sequence.

Atomic multiplet scattering simulations including charge transfer (CT) effects were done using charge transfer ligand field multiplet theory, implemented in the CTM4XAS software(*35*). Atomic multiplet simulations in Fig. 1g for $Ti^{4+}$ ions were performed in D4h symmetry and included a charge transfer (CT) term to account for the hybridization between Ti-3*d* and O-2*p* states in the $TiO_6$ cluster, i.e. a $3d^0+3d^1\underline{L}$ configuration, where $\underline{L}$ indicates an hole in the O-2*p* band. The CT is included in an Anderson impurity type of model where two parameters are necessary to describe the system: the energy difference between the Ti-*3d* and the ligand band (called $\Delta$ here), and the difference $U_{pd}$-$\underline{U}_{dd}$ between the core hole potential ($U_{pd}$) and the Hubbard U value ($U_{dd}$) for Ti-d electrons. Here we used, $\Delta$=5 eV and $U_{pd}$-$\underline{U}_{dd}$ = 2eV. Finally, the transfer integrals used were 2 and 1 for $b_1$, $a_1$ ($d_{x2-y2}$, $d_{z2}$) and $b_2$, e ($d_{xy}$, $d_{xz}/d_{yz}$) orbitals, respectively. This choice is commonly used for octahedral or close to octahedral systems and the reasoning behind is that the probability for electron transfer is higher (in this case two times higher) for orbitals pointing in the bond direction, as $d_{x2-y2}$, $d_{z2}$. The Slater Integrals were reduced by 90% respect the corrected Hartree-Fock values. The broadening used were 0.1 eV for the Gaussian broadening, to account for the energy resolution, and different Lorentzian broadening for each of the 4 peaks of the spectrum, namely half width at half maximum of 0.125 eV, 0.33 eV, 0.55 eV, and 0.7 eV for the four XAS peaks (from the first to the last). In the calculations of the Ti-3*d* XLD as function of the polarization we changed only the crystal field parameters, which were $10D_q$=2.15, and $D_s$=-0.020, $D_t$=-0.000 for $P_{up}$ and $D_s$=-0.023, $D_t$=-0.001 for $P_{dn}$.

First-principles simulations were performed with Density Functional Theory (DFT) using the VASP package (*36*, *37*). In order to better cancel the spurious self-interactions errors inherent to DFT, we have used the meta-GGA SCAN functional (*38*) that was previously shown to be suited for capturing the physical properties of transition metal perovskite oxides with a 3*d* element, even without any U parameter on 3*d* states (*39*). The choice of this functional is also appealing since Ti 3*d* occupancies may vary when moving away to the interface with LAO. While DFT+U method may require different U potential depending on Ti $d$ state occupancy and formal oxidation state – U is a potential acting on a subset of orbitals and correcting delocalization errors—, the SCAN-no-U functional may provide a fair description of the physical effects through the different layers upon adding Ti $d$ electrons as it captures key properties of $d^1$ titanates (*39*) and $d^0$ titanates (*40*). Nevertheless, the SCAN functional can sufficiently amend delocalization errors for 4$f$ states and we have added a U potential of 5 eV on Eu 4$f$ states in order to capture the insulating character of EuTiO$_3$ in bulk. It yields a band gap of 0.53 eV and a 4$f$ band just below the Ti 3$d$ states as found in more sophisticated but computationally more demanding hybrid functional DFT (*18*). Projector Augmented-Wave (PAW, Ref. (*41*)) potentials are used with the following valence configurations for the different ions: $4s^2 3d^2$ (Ti), $3s^2 3p^1$ (Al), $4s^2 4p^6 5s^2$ (Sr), $5s^2 5p^6 6s^2 5d^1$ (La), $4s^2 4p^6 5s^2 4f^7$ (Eu), $2s^2 2p^4$ (O). The total energy is converged with a cut-off of 500 eV, accompanied with a 6x6x1 Monkhorst-Pack k-mesh (*42*). The slab configuration consists of a (STO)$_{10}$/(ETO)$_2$/(LAO)$_6$ tricomponent material with 20 Å of vacuum on top in order to avoid interactions between periodic replica of the slabs. In order to mimic the spontaneous polarization of the Ca-STO, we have applied a small compressive strain to STO (1.6%) that yields a polarization of 18 μC.cm$^{-2}$. The slab is then initialized with a polar STO and we impose that the first 4 unit cells at the bottom are pinned during the whole calculation while all other atomic positions are optimized until forces acting on each atom are lower than 0.01 eV/ Å. A ferromagnetic order is imposed to Eu$^{2+}$ spins throughout the whole calculations (consistent with the ground state of metallic ETO(*18*)); no initial spins are imposed on Ti cations.

**Supplementary Material**

**Fig. S1**

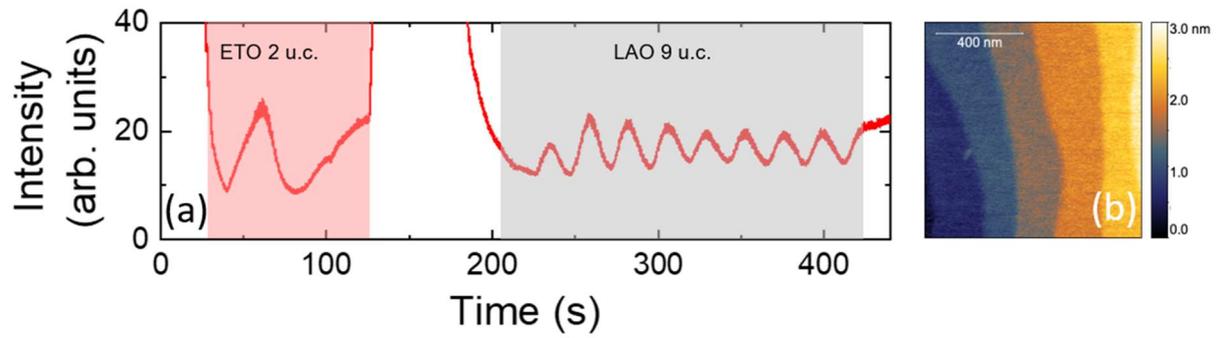

(a) RHEED intensity as function of time during the preparation of a LAO/ETO//Ca-STO heterostructure. (b) AFM image of the sample after growth. Each terrace is separated from its neighbours by a 4 Å high step.

**Fig. S2.**

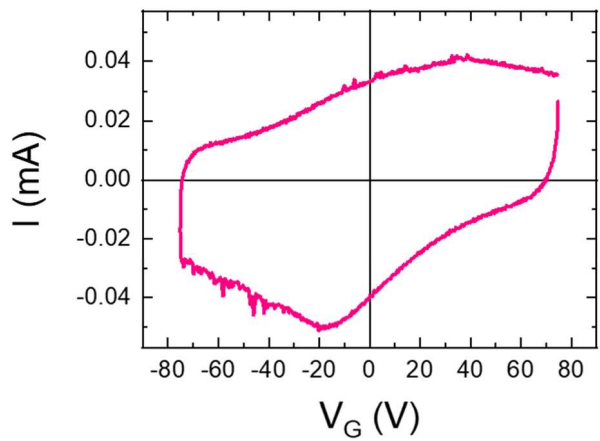

Current vs voltage loop at 2 K. The shape of the loop differs from what can be measured in a classical ferroelectric like PZT or BiFeO$_3$ because here the dielectric constant is huge (>10000 vs <100 is typical ferroelectrics) and the polarization is small (~1 µC/cm² vs 50-100 µC/cm² in typical ferroelectrics). This implies that the background signal, associated with the dielectric contribution, is large and the peaks, associated with polarization switching, are less pronounced. Yet, the integration of this signal does produce a polarization loop, cf. Fig. 3b.

**Fig. S3.**

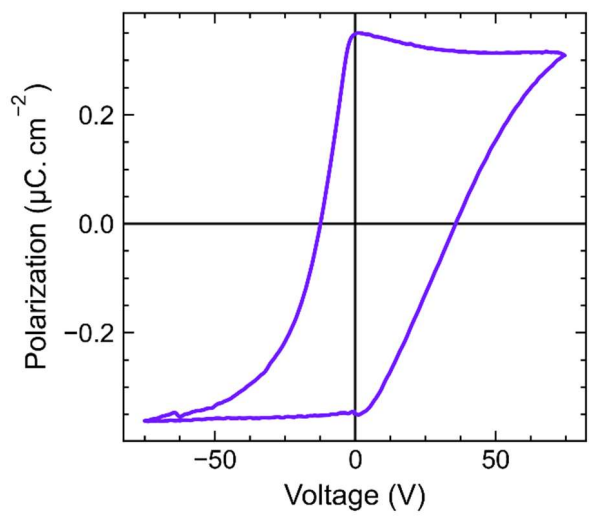

Remanent polarization loop at 2 K measured with a voltage sequence equivalent to "PUND" measurements. The dielectric contribution visible in Fig. 3b is then absent and only the switched polarization contributes.

**Fig. S4.**

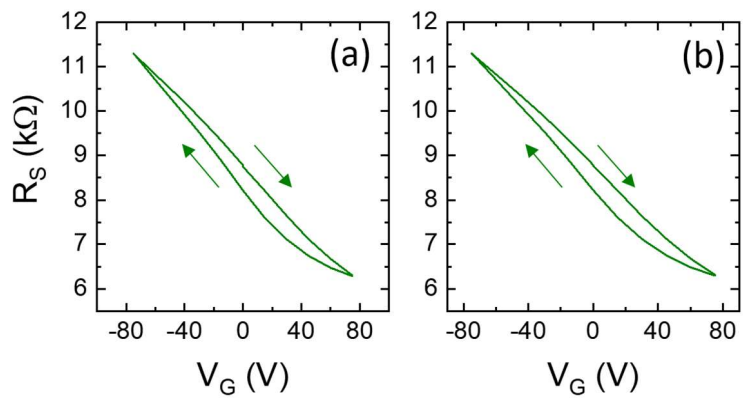

Two R vs $V_G$ cycles as in Fig. 3b measured consecutively showing a clockwise behavior. They are virtually identical, evidencing that the resistance is modulated by ferroelectricity rather than by irreversible effects occurring on the first cycle in standard STO 2DEGs which is anticlockwise. Subsequent gate sweeps in such standard STO 2DEGs are then anhysteretic (*27*).

**Fig. S5.**

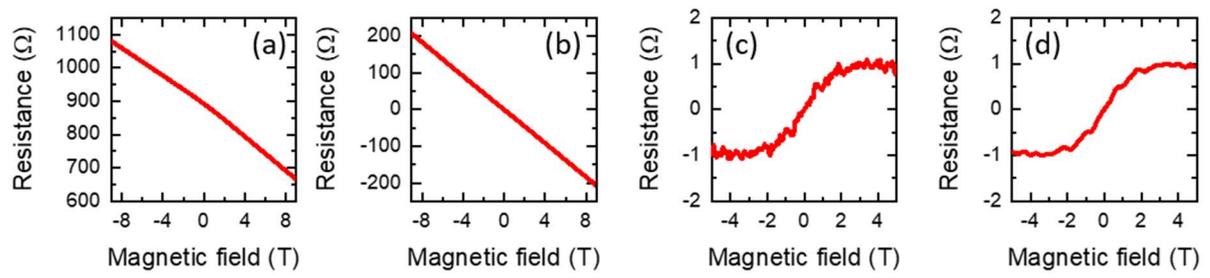

(a) Raw transverse resistance measured at 6 K in the $P_{up}$ state. (b) Same data after antisymmetrization.

(c) AHE after the subtraction of a linear slope. (d) Same as in (c) after smoothing.

**Fig. S6.**

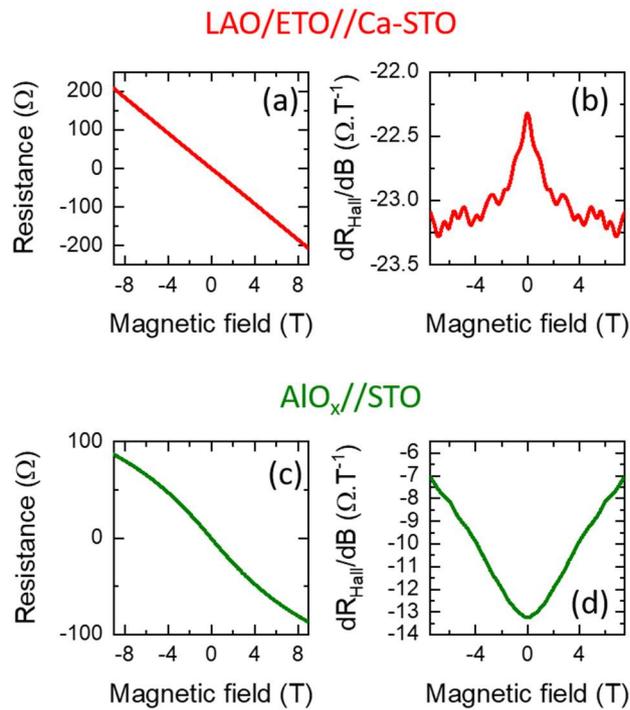

Hall effect (a) and field derivative of the Hall effect (b) for a LAO/ETO//Ca-STO sample. Hall effect (c) and field derivative of the Hall effect (d) for a AlOx//STO sample. The inverted-V shape and V shape of the data in (b) and (d), respectively, indicate a non-linear component in the Hall signal, but with a sign opposite in LAO/ETO//Ca-STO compared to AlOx//STO. While in AlOx//STO (and in standard LAO/STO 2DEGs) this non-linearity is well explained by standard two-electronic bands contributions, the non-linear component in LAO/ETO//Ca-STO has a different origin, only visible when a magnetic layer is present(*19*), and thus ascribed to anomalous Hall effect.

**Fig. S7.**

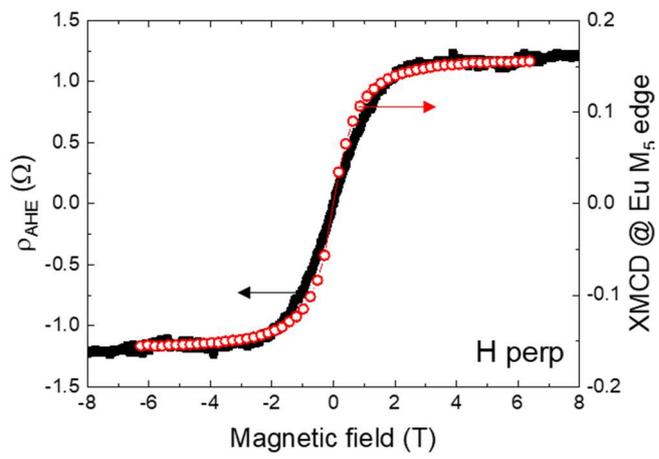

Comparison of the field dependence of the AHE (left axis) and the XMCD at the Eu $M_5$ edge (right axis), both measured at 2 K and with the magnetic field perpendicular to the plane.

**Fig. S8.**

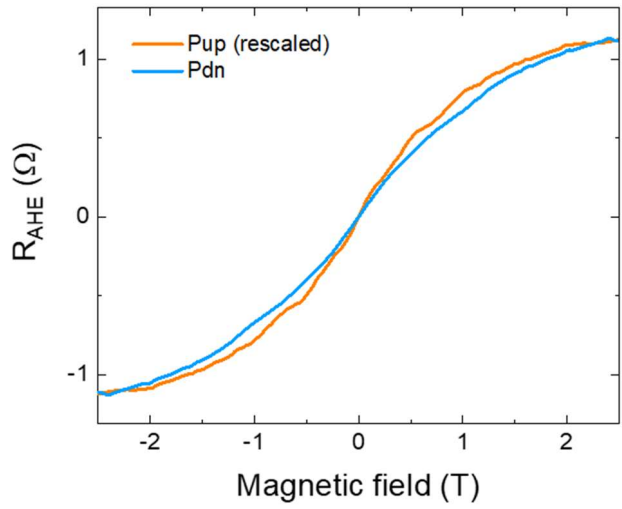

Low magnetic field comparison of the AHE signal at 2 K for both remanent polarization state, after rescaling the $P_{up}$ signal to match the amplitude of the $P_{dn}$ signal at high magnetic field.

**Fig. S9.**

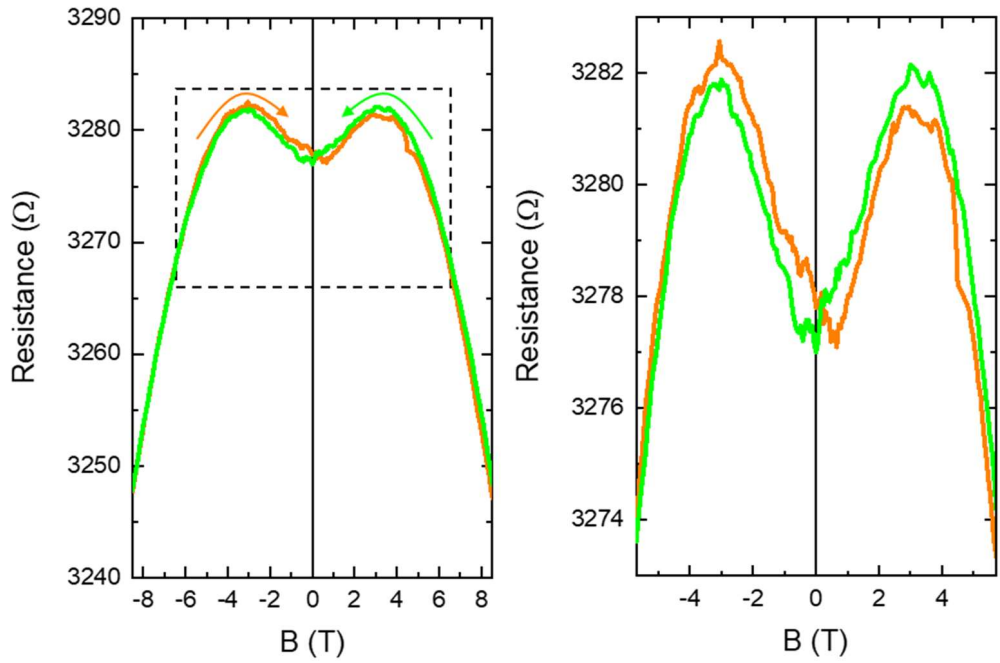

MR measured with the magnetic field in the plane at 2 K. The right graph is a zoom of the region highlighted in the left graph. The coercive fields are in the 3-5 kOe range, which compares well with the coercive field from Ti XMCD (Fig. 2e) measured on a different sample, that are around 2 kOe.

**Additional references**

34. C. Piamonteze, U. Flechsig, S. Rusponi, J. Dreiser, J. Heidler, M. Schmidt, R. Wetter, M. Calvi, T. Schmidt, H. Pruchova, J. Krempasky, C. Quitmann, H. Brune, F. Nolting, X-Treme beamline at SLS: X-ray magnetic circular and linear dichroism at high field and low temperature. *J Synchrotron Rad*. **19**, 661–674 (2012).

35. E. Stavitski, The CTM4XAS program for EELS and XAS spectral shape analysis of transition metal L edges. *Micron*. **41**, 687–694 (2010).

36. G. Kresse, J. Hafner, *Ab initio* molecular dynamics for liquid metals. *Phys. Rev. B*. **47**, 558–561 (1993).

37. G. Kresse, J. Furthmüller, Efficiency of ab-initio total energy calculations for metals and semiconductors using a plane-wave basis set. *Computational Materials Science*. **6**, 15–50 (1996).

38. J. Sun, A. Ruzsinszky, J. P. Perdew, Strongly Constrained and Appropriately Normed Semilocal Density Functional. *Phys. Rev. Lett.* **115**, 036402 (2015).

39. J. Varignon, M. Bibes, A. Zunger, Mott gapping in 3 d A B O 3 perovskites without Mott-Hubbard interelectronic repulsion energy U. *Phys. Rev. B*. **100**, 035119 (2019).

40. A. Paul, J. Sun, J. P. Perdew, U. V. Waghmare, Accuracy of first-principles interatomic interactions and predictions of ferroelectric phase transitions in perovskite oxides: Energy functional and effective Hamiltonian. *Phys. Rev. B*. **95**, 054111 (2017).

41. P. E. Blöchl, Projector augmented-wave method. *Phys. Rev. B*. **50**, 17953–17979 (1994).

42. H. J. Monkhorst, J. D. Pack, Special points for Brillouin-zone integrations. *Phys. Rev. B*. **13**, 5188–5192 (1976).